\documentclass[sigconf, nonacm]{acmart}
\AtBeginDocument{%
  }

\usepackage{amsmath}   % in your preamble
\begin{document}

\title{Reading the Neighborhood: Text-Attentive Graph Reranking for Automated Subject Indexing}

\author{Lakshmi Rajendram Bashyam}
\email{l.rajendram-bashyam@zbw.eu}
\orcid{0009-0009-9455-2829}
\affiliation{%
    \institution{ZBW -- Leibniz Information Centre for Economics}
    \city{Kiel} 
    \country{Germany}
}

\author{Ralf Krestel}
\email{rkr@informatik.uni-kiel.de}
\orcid{0000-0002-5036-8589}
\affiliation{%
    \institution{ZBW -- Leibniz Information Centre for Economics \\
    \& Kiel University}
    \city{Kiel} 
    \country{Germany}
}

% \renewcommand{\shortauthors}{Bashyam et al.}

\begin{abstract}
 
Subject indexing assigns descriptors from a controlled vocabulary to bibliographic records, a task characterized by thousands of labels, few positive examples per label, and a long tail of rarely used descriptors. Existing extreme multi-label classifiers treat the vocabulary as a flat list, ignoring the semantic relations that thesauri encode.

We present TAGHN, a heterogeneous graph reranker over a per-document subgraph of retrieved candidates augmented with the thesaurus neighborhood. The relevance of a candidate descriptor is estimated jointly from its similarity with the document and also the degree to which the candidate's thesaurus neighborhood matches the document content.  A decomposed scoring head combines interpretable graph-derived features with retrieval signals. 

On STW, an economic thesaurus of 6,198 descriptors, TAGHN reaches 0.380 macro F1 against 0.346 for a fine-tuned cross-encoder and 0.278 for a fixed-vocabulary XMLC classifier. In terms of micro F1, it outperforms reranking models and stays competitive (-3.0) with the best fixed-vocabulary XMLC model. Qualitative analysis shows the graph resolving errors that per-label scoring cannot, selecting a narrower gold descriptor where an unstructured reranker settles for its broader parent.
\end{abstract}




\maketitle

\section{Introduction}
Subject indexing is the task of enriching a metadata record for a textual resource (a document representing an article or a book, for example) with multiple subjects that cover the content of the resource. This is typically done by professional subject indexers who are also domain experts, i.e., librarians working at the library that has the resource in its collection. In order to benefit the downstream use cases of library subject indexing, descriptors are taken from a pre-defined, controlled vocabulary, such as a thesaurus.

However, with the growing number of library catalogs, it is challenging to keep pace purely with manual workflows. Using computational methods, typically machine learning methods, to automatically identify appropriate subject terms from a controlled vocabulary is called automated subject indexing. It allows libraries to keep pace with the growing volume of digital materials and ensures that their collections remain organized and easily navigable in the digital age.

The size of the thesaurus used for indexing purposes can vary significantly across different institutions. However, many of the thesaurus maintained by the libraries have several thousand descriptors ranging from 1,000-20,000+ subjects. A large label set combined with limited training data per label makes the task an extreme multi-label classification (XMLC) problem.. XMLC tasks are also usually characterized by highly skewed label distribution with long tail labels. This phenomenon is also observed in automated subject indexing. Depending on the amount of training data available, it can also be considered a low-resource setting. 

Adapting a strategy to suit this need involves identifying where traditional approaches fail. Flat XMLC classifier may address the typical XMLC problem of handling a large label set, but it treats each subject independently, and the label quality is bound to the strength of the training signal. Thus, the performance of such models is competitive for head labels but suffers for the long tail which only appears a few times in the training data. They also tend to fail on unseen labels because the model cannot predict a label that it has never seen before. 

Retrieval-based rerankers such as cross-encoders are not bound by the label frequency constraint since they predict a subject by evaluating each document- subject pair independently. The subjects are assigned without any knowledge of where they are positioned in the thesaurus. However, subject indexers are advised to be as specific as possible (i.e., pick subjects that are closer to the leaves of the descriptor hierarchy), but if a document treats very broad topics, then the content can only be described by more generic descriptors. Choosing an adequate number of descriptors and the correct level of granularity and specificity lies within the responsibility and discretion of the subject indexing expert. Thus, the reranker cannot recognize that a more specific descriptor in the same branch of the hierarchy fits the document better. For example,  a document about ``Automation, Innovation and Economic Crisis: Surviving the Fourth Industrial Revolution`` has a gold subject: Technological change. It doesn't have any lexical anchor from the document text. However, it is reachable from Automation and Innovation which are present in the document text.

This is the observation we build on: a subject's relevance to a document is evidenced by whether its neighborhood matches that document too, and the same descriptor should be read differently for different documents. The neighborhood of a subject in the thesaurus is defined by its broader parent, narrower children, and related terms.

We cast automated subject indexing as a two-stage retrieve-then-rerank problem and present \textbf{TAGHN}\footnote{Code and configuration: \url{https://github.com/Lakshmi-bashyam/taghn-si.git}}, a relational reranker that scores each candidate descriptor over a per-document subgraph of retrieved candidates augmented with non-assignable BT/NT/RT context nodes from the thesaurus. To our knowledge, this is the first reranker to condition message passing on the document, so that the same descriptor is read differently depending on the text being indexed
\begin{itemize}

\item We introduce a document-conditioned message-passing layer with two multiplicative channels: a per-relation edge-type gate and a three-way attention over target, neighbor, and document.
\item We fold a document–descriptor cross-encoder into the same attention through a gate initialized at zero, so the model begins training without the cross-encoder signal and learns how much of it to use.
\item We design a decomposed scoring head that combines interpretable graph-derived features (semantic, NT-, and RT-match) with retrieval signals, and a specificity loss that encodes the indexing convention of preferring a narrower descriptor over its broader parent.
\item We evaluate TAGHN on a bilingual thesaurus of 6{,}198 descriptors, with a per-frequency-bin analysis across the head and tail and a qualitative study of how the graph resolves errors that per-label scoring cannot.
\end{itemize}

\section{Related work}
Automated subject indexing assigns a set of descriptors from a controlled vocabulary to a bibliographic record. The earliest indexing models were lexical models that match document text with label names. Associative models learn the correlation of labels in the document text. Annif \cite{annif} is a popular CLI-based toolkit that consolidates several models, including lexical models such as tfidf, mllm and associative models like fasttext \cite{ft}, omikuji \cite{prabhu2018parabel}\cite{khandagale2020bonsai}. It provides various methods to ensemble these models together, including a neural network variant. It has become a de facto reference point for library indexing across vocabularies and languages. Concept drift analysis on short text \cite{conceptdrift} for the task of automated subject indexing also found that fusion approaches outperformed other methods. However, all of these methods consider subjects as a large label list and do not take into account the structural relation between the subject from a thesaurus.

The large number of subjects in the thesaurus inflates the label list to be considered for each document. Thus, subject indexing is most often cast as an XMLC task. Models that are designed to handle the challenges of XMLC are either a one-vs-all classifier, such as DiSMEC\cite{babbar2017dismec}, Parabel\cite{prabhu2018parabel} and its variant bonsai\cite{khandagale2020bonsai}. These are tree-based methods that reduce the label space through label co-occurrences. Attention-based methods such as AttentionXML \cite{you2019attentionxml} and transformer-based models such as LightXML\cite{jiang2021lightxml} and X-Bert\cite{xbert} provided better text representation and thereby better performance. A more recent model,  the XRTransformer\cite{xrtransformer,yu2022pecos} uses transformer based model that operates in three stages, first clustering the label space to reduce the label size, followed by a matcher module that identifies the correct cluster, and a rerank step to rescore the labels inside the cluster. This restricts the labels in the retrieved clusters, thus affecting the label coverage. All the aforementioned models offer competitive performance on head labels but fail to generalize to tail labels, especially with 0 scores for unseen labels.

Models that exploit dual-encoders such as DEXML\cite{dexml} and retrieval-augmented encoders such as RAE-XMC\cite{rae} target the tail by embedding inputs and label text in a shared space, trading the strong head-label training signals for tail generalisation. Our Stage-1 adopts this retrieval view to maximize recall, but rather than commit to a single OVA/DE trade-off, we defer the precision decision to a structure-aware second-stage reranker.

Retrieve-rerank pipelines shortlist a comparatively larger list of candidates than the result set with an efficient retriever and defer to a more complex model to rescore them. Multi-vector and dense retrievers-ColBERT\cite{khattab2020colbert}-style late interaction and the unified dense/sparse/multi-vector encoding of BGE-M3\cite{chen2024bge} supply strong first-stage recall, while LLM-based cross-encoders that read query and candidate jointly provide accurate but expensive reranking. Infer-retrieve-rank \cite{irerera} model performs the XMLC task within the same pipeline. It generates labels in a few-shot setting using frozen LLMs and a retriever that reranks the result. TAGHN follows this two-stage process, but its reranker does not score candidates independently. It scores them jointly over a per-document subgraph.

Message passing over a graph was introduced with Graph Neural Network (GNN). Graph convolution neural network\cite{gcn} aggregated its neighbors with a fixed weight. GAT\cite{gat} extended this idea with a content-dependent attention on the neighbors. GATv2\cite{gatv2} makes that attention dynamic rather than static-the variant adopted by TAGHN's node-level channel B. Multi-relational heterogeneous graphs are effectively processed by R-GCN\cite{rgcn}, which learns a separate transformation per relation type. HGT\cite{hgt} extends it to include node types by parameterizing attention between node and relation type. TAGHN draws these ingredients together, but conditions all of them on the document. Channel A gates each relation type with a document-dependent scalar, and Channel B computes node-level attention jointly from the target node, the neighbor, and the document embedding. While  R-GCN and HGT learn type parameters that remain fixed at inference, TAGHN recomputes its per-relation gates and attention weights for every document.

A parallel line of work injects label dependencies into classification. In XMLC, ECLARE\cite{eclare}  and GalaXC\cite{saini2021galaxc} augment label representations with correlation graphs built from co-occurrence statistics, and BGNN-XML\cite{zong2022bgnn} extends this to bilateral text–label graphs. In hierarchical text classification, HiAGM\cite{hiagm}  and HGCLR\cite{hgclr} encode a global taxonomy with graph or Graphormer\cite{graphormer} encoders and fuse it with the text representation. Methods like K-HTC\cite{khtc} incorporate label hierarchy into hierarchical classification by constructing a graph-based encoder over the label taxonomy and a contrastive-learning objective that aligns text representations with label representations. 

\section{Problem formulation}

\subsection{Task: subject indexing}
Subject indexing in libraries is usually done by referring to a controlled vocabulary, i.e., a set of established authority descriptors. We use the STW Thesaurus for Economics\footnote{STW Thesaurus for Economics, \url{https://zbw.eu/stw}}, which is developed and maintained in-house at ZBW -- Leibniz Information Centre for Economics\footnote{ZBW -- Leibniz Information Centre for Economics, \url{https://www.zbw.eu}}. It contains around 6,000 descriptors that are categorized into seven subthesauri and linked by broader/narrower relationships in a polyhierarchy. The thesaurus is updated every year by a board of subject indexing experts and published on the web as linked open data in SKOS. Librarians tasked with subject indexing will identify and assign matching descriptors from the thesaurus to a publication first, and then possibly cover remaining salient aspects with free keywords, which will become candidates for additions to the next edition of the thesaurus later on.

While some thesauri contain a number of descriptors in the order of 20,000 or more, the label set size for the thesaurus used in this work contains a  little over 6,000 labels. This in combination with the variable number of labels for each record still qualifies our subject indexing use case comfortably as a so-called extreme multi-label text classification (XMLC) task. XMLC is also characterized by the existence of ``long-tail`` labels, i.e., labels that do not occur very frequently in the available training data. A few ``head`` labels appear on the majority of the documents, but the vast majority are ``tail`` labels that might only appear once or twice in the entire training data.

Traditional text classification treats labels as independent strings and has no notion of how two labels, e.g., construction cost and real estate, relate to each other. Subject indexing, however, operates on taxonomy/ontology. The labels are not independent; they are deeply interconnected concepts governed by strict semantic rules.

\subsection{The thesaurus as a heterogeneous graph}
We assign descriptors as subjects for the document. These descriptors are taken from the STW, which covers economics along with its neighboring business and social-science disciplines. The thesaurus is published as linked open data using the SKOS vocabulary. Each descriptor is a  \texttt{skos:Concept} and the relationship between the descriptors is expressed by  SKOS semantic properties. Exploiting SKOS relationships gives us the ability to treat the labels as a structured graph and not just a flat label list. 

We represent the thesaurus as a directed edge-typed graph $G = (L, E)$, where $L = \{\ell_1, \dots, \ell_{|L|}\}$ is the set of descriptors (labels) and $E$ is a set of typed edges. 
Each edge is a triple

\begin{equation}
  (\ell_i, \ell_j, r), \qquad r \in \mathcal{R} = \{\textsc{bt}, \textsc{nt}, \textsc{rt}\},
\end{equation}

where $\ell_i$ is the source concept, $\ell_j$ the target concept, and $r$ the relation type connecting them.

The triple $(\ell_i, \ell_j, \textsc{bt})$ denotes that the $\ell_j$ is a more general parent to $\ell_i$. In other words, it is the hierarchical parent of $\ell_i$. Similarly, triple $(\ell_i, \ell_j, \textsc{nt})$ denotes that the $\ell_j$ is a narrower and a specific term to $\ell_i$. The \textit{BT} and \textit{NT} relation types are strict inverses of each other. Modeling them as separate relations better captures the directional signal between the labels than representing the relationship as an undirected link.  
The triple $(\ell_i, \ell_j, \textsc{rt})$ denotes a related edge between labels $\ell_j$ and $\ell_i$. While the BT/NT edge induces hierarchy in the thesaurus, the related edge represents an associative, otherwise unspecified, semantic relationship between the labels. The three different types of edges make the graph heterogeneous in nature.

 Beyond relations on the descriptor level, our thesaurus also has a coarser, superordinate classificatory layer. The descriptors are grouped into a set of subject categories, which in turn are organized in seven sub-thesauri. The subject categories are arranged in a hierarchical structure with BT, NT relation edges between them as well and are therefore also part of the graph. Following the broader-term links upward from any descriptor eventually reaches a top-level category. This means every descriptor can be seen to naturally inherit semantic context from its parent category. And since the thesaurus is a poly-hierarchy, any descriptor with an overlapping meaning can belong to more than one subject category and therefore be connected to more than one of the seven sub-thesauri at once. We use this membership assignment as a signal

In addition to the structural metadata of label relation edges, subjects that are part of the thesaurus are also represented by alternate labels (synonyms) and hidden labels. Some of the labels also contain descriptions about themselves and scope notes on how it has to be assigned. The label statistics are given in the table \ref{tab:metadata-stats}
\begin{table}[t]
\centering
\caption{Label metadata and relation coverage across the $6,198$ active descriptors.}
\label{tab:metadata-stats}
\begin{tabular}{lrr}
\toprule
Field & Count & \% \\
\midrule
prefLabel EN (label)                 & 6{,}198 & 100.0 \\
prefLabel DE                         & 6{,}198 & 100.0 \\
\textbf{Any altLabel (EN+DE+narrower+related)} & \textbf{5{,}136} & \textbf{82.9} \\
\quad altLabel EN                    & 3{,}458 & 55.8 \\
\quad altLabel DE                    & 4{,}538 & 73.2 \\
\quad altLabelNarrower       & 426     & 6.9  \\
\quad altLabelRelated        & 213     & 3.4  \\
scopeNote EN                         & 523     & 8.4  \\
Broader descriptors                  & 4{,}310 & 69.5 \\
Narrower descriptors (inverse)       & 1{,}751 & 28.3 \\
Related descriptors                  & 5{,}214 & 84.1 \\
Thesaurus category ids                   & 6{,}198 & 100.0 \\
\bottomrule
\end{tabular}
\end{table}

Each document text consists of a title and an abstract concatenated. The document text is represented as a 1024-dimensional vector using a single frozen domain-adapted BGE-M3\cite{bge-reranker}\cite{chen2024bge} model. Similarly, the subject categories are encoded into the same embedding space using their preferred labels. For descriptors, the label text consists of the preferred, alternative, and hidden label text. It is concatenated with definition and scope notes whenever available. This enriches the label representation while projecting it onto the embedding space. 
Because this encoder is trained contrastively so that semantically related texts are close to each other, cosine similarity in this space approximates semantic relatedness for both document–concept and concept–concept pairs. This shared space is what allows a document to be mapped directly against individual candidate descriptors.

\subsection{Retrieve-and-rerank}

Training an ML model for an XMLC task can be challenging. Finetuning the model on the entire label set can work well for head labels (labels that are frequently observed in the training dataset). But it can negatively impact tail labels (labels that are hardly observed in the train dataset) since they only contribute a very weak signal for the model. One of the ways to effectively capture tail labels is to design a system that does not depend on the label frequency. A retrieve-and-rerank pipeline is one of the most effective ways to handle XMLC tasks without under performing on the tail labels. 

The architecture of our system is divided into two stages. Stage-1 is the candidate generation stage of our pipeline.  For a document (represented by the metadata elements title and abstract), the objective of Stage-1 is to select a small set from thousands of subjects present in the thesaurus, which the downstream module then reranks. This is a recall-maximizing stage that does not blow up the candidate set size excessively. 
Formally, for each document we seek candidates $C$ such that the recall $|C \cap G| / |G|$ (with $G$ the gold labels) is as high as possible under a budget $|C| \approx$ a few hundred. The gold labels that are not in the candidate list cannot be recovered later in the pipeline; hence, this creates an upper ceiling for the recall. The precision is left to be improved in the second stage.

Rather than scaling up a single retriever, we combine four complementary sources. A dense multi-vector BGE-M3\cite{bge-reranker}\cite{chen2024bge} retriever (dense, sparse, and ColBERT\cite{khattab2020colbert} max-similarity) supplies semantic matches. A trained subject category classifier restricts attention to the relevant thesaurus branches. An NER-based mapper links domain entities extracted from the text to descriptors. Finally, a surface exact-match over labels and alternative labels safeguards rare tail terms, which often appear verbatim.

Each source $s$ returns its own top-$k_s$ candidates $C_s$, and we take the union $C = \bigcup_s C_s$. Because these signals capture different kinds of relevance, their combination lifts recall well beyond any single retriever, while the per-source cutoffs $k_s$ keep $|C|$ compact.

\begin{figure}
    \centering
    \includegraphics[width=\linewidth]{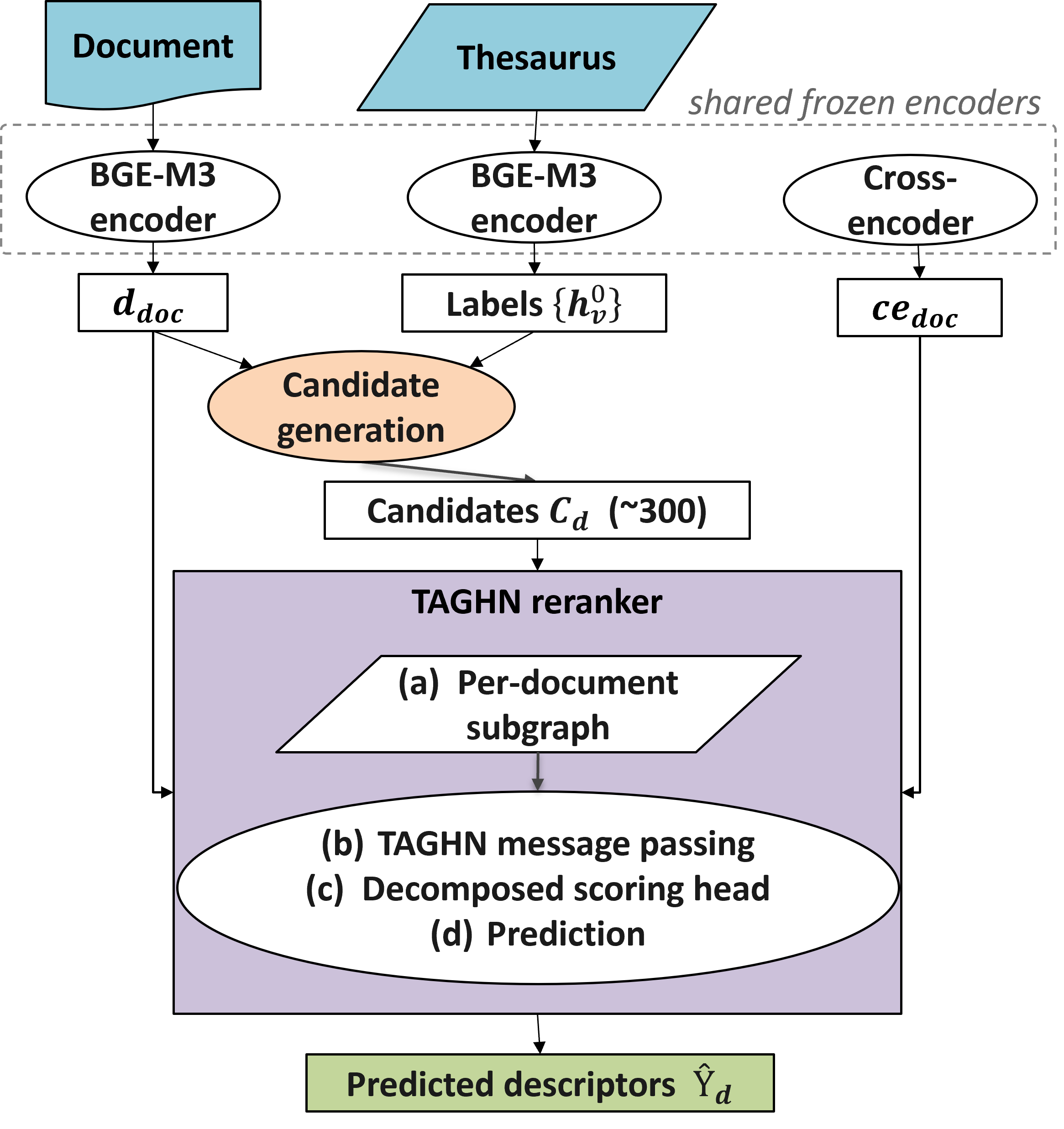}
    \caption{End-to-end architecture of \textsc{TAGHN}.
    \Description{Flowchart of the TAGHN pipeline. A document and the thesaurus are encoded by frozen BGE-M3 encoders and a cross-encoder. Stage-1 generates about 300 candidate descriptors, which the TAGHN reranker processes through a per-document subgraph, message passing, a decomposed scoring head and prediction, producing the predicted descriptors.}
    % A document and the thesaurus are encoded by a shared frozen BGE-M3 model; Stage~1 retrieves candidate descriptors, and Stage~2 refines them with document- and cross-encoder-conditioned relational message passing, a decomposed interpretable scoring head, and threshold-based prediction with BT-suppression.
    }
    \label{fig:placeholder}
\end{figure}

Stage-1 narrows our nearly 6,000 subject vocabulary to candidates of $C_d \subseteq \mathcal{L}$ of ~300 subjects. However, Stage-1 scores candidates independently of each other without any awareness of each other or how they relate within the thesaurus. Stage-2 takes these candidates as inputs and reranks them to identify the correct candidate subjects. Stage-2 is essentially a scoring function $f_\theta$ that assigns a score to each candidate $c \in C_d$, not just grounded on the document $d$ but also based on how they interact with other candidates. The scores are calibrated based on the candidate neighborhood, which is defined by narrow (NT), broader (BT), and related (RT) subject nodes.  Our reranker $f_\theta$ is TAGHN, a relational graph network over the per-document subgraph. The architecture of the TAGHN is given in Figure \ref{fig:placeholder}.

Candidates enter Stage-2 with scores and evidence accumulated from Stage-1 summarized in Table~\ref{tab:stage1-features} as a feature vector populated from the BGE-M3 embedding matching step, LLM-based NER matching step, and the subject category identification step, followed by feature flags on how many candidates were generated and candidate source in Stage-1. Along with these features, each candidate is represented by the label embedding $\mathbf{h}^c_0 \in \mathbb{R}^{1024}$. Whereas the document $d$ is represented by $\mathbf{d}_{\text{doc}} \in \mathbb{R}^{1024}$
that was derived from frozen BGE-M3 dense representation $\mathbf{r}_c$ of candidate label and document text. 

Given the candidate scores, the final assignment applies a learned decision threshold $\theta$,
\begin{equation}
\hat{Y}_d = \{\, c \in C_d : s_c \ge \theta \,\}
\end{equation}

The reranker is trained to maximize agreement between $\hat{Y}_d$ and the gold set $Y_d$, optimising a relevance objective augmented with auxiliary terms that exploit the thesaurus structure. We evaluate with set-based metrics (precision, recall, $F_1$).

\section{TAGHN reranker}

TAGHN reranks the Stage-1 candidates set jointly over 

\begin{itemize}
    \item Semantic match between each subject and the document
    \item Structural positioning of each subject and how it relates to it's neighborhood subject nodes.
\end{itemize}

Given a document $d$ with embedding $\mathbf{d}_{\text{doc}}\in\mathbb{R}^{1024}$ and its candidate set $C_d$, the Stage-2 module 
\begin{itemize}
    \item Constructs a per-document sub-graph that includes the candidate set from Stage-1, enriched with context from thesaurus
    \item Apply two layers of message passing over the label sub-graph to obtain a structurally aware label representation
    \item Along with the attention weights from the message passing layers, seven interpretable features from Stage-1 are scored by an MLP - decomposed scoring head
    \item Converts score to a subject set through a learned threshold, followed by a hierarchical suppression 
  \end{itemize}
The entire model is trained end-to-end with a relevance objective augmented by structural losses. The important design choice of TAGHN is that every part of the reranker, including the subject node and its relevance to its neighbors, is conditioned on the document $\mathbf{d}_{\text{doc}}$. This ensures that the subject representation is computed differently based on the document text and can be considered a discriminatory signal. 

\subsection{Subgraph construction}
A subgraph is constructed from the candidate subject set based on the relationship edges between them. The subgraph is also augmented with context nodes (nodes that add structural information but are not part of the candidate set) based on the thesaurus neighborhood of the subject. A per-document subgraph is constructed based on
\begin{itemize}
    \item BT (Broader) parent of each candidate (one per candidate, when it exists)
\item top-$K_{\text{NT}}{=}5$ NT (Narrow) children of each candidate;
\item top-$K_{\text{RT}}{=}3$ RT (Related) neighbors of each candidate.
\end{itemize}

The table \ref{tab:metadata-stats} shows the number of broader, narrower, and related descriptors. The more general the subject, the higher the number of narrow and related subjects it has. Hence, we only consider the top \textit{k} descriptors ranked by $\cos(\mathbf{h}^0_v, \mathbf{d}_{\text{doc}})$ on their initial label embedding. It ensures that a single candidate does not overcrowd the subgraph. 
The model's core intuition is that a subject is more likely to be relevant to a document based on whether its neighborhood subjects match the document. For example, a broad term like labor market policy can gather evidence from its narrower children (minimum wage, unemployment insurance, \ldots{}) and decide whether they, too, look relevant to this particular document.

Every node in the graph is a subject and is initialized with embedding $\mathbf{h}^0_v$ from a frozen, domain adapted bge-m3 model. The edges denote the relation between the subject, which is of type BT, NT: which is an inverse of BT and RT: Which is a symmetric relation. The context nodes are not part of the Stage-1 candidate subset and are marked by $\text{ctx}(v)\in{0,1}$ flag. In addition, a self-loop is added to every node to ensure that the subject relevance to the document is retained as part of the label representation. 

\subsection{Document-conditioned message passing layer}
A layer updates each node in the graph by aggregating its neighbors through two document-conditioned channels 
\subsubsection{Channel A}: Edge type gate 
Information populated from a relation can vary based on document content. We therefore compute a document-dependent gate for each relation type
\begin{equation}
\boldsymbol\beta = \sigma\!\big(W_g\,\mathbf{d}_{\text{doc}}\big) \in (0,1)^{4},
\end{equation}

where $W_g$ is a learned projection and $\sigma$ is the logistic sigmoid, which restricts each gate to $(0,1)$. Each component $\beta_r$ scales the entire contribution of relation type $r$. 

\subsubsection{Channel B}: Neighbor level attention 
Within each relation type, a neighbor can have a varied contribution to the node. For a directed edge from the neighbor $j$ to the node $i$ of type $r$, an attention logit is computed jointly from the target node, the source neighbor, and the document:
\begin{equation}
e^{r}_{ij} = \mathbf{a}_r^{\!\top}\mathrm{LeakyReLU}\!\big(W^Q_r\mathbf{h}_i + W^K_r\mathbf{h}_j + W^D_r\mathbf{d}_{\text{doc}}\big)
\end{equation}

with per-relation query $W^Q_r$, key $W^K_r$, and document $W^D_r$ projections attention vector $\mathbf{a}_r$, and a LeakyReLU non-linearity (negative slope $0.2$). For relational edge types, logits are normalized across each target's incoming neighbors of that type by a softmax, $\alpha^{r}_{ij} = \mathrm{softmax}_{j \in \mathcal{N}_r(i)}, e^{r}_{ij}$, resulting in attention weights that sum to one per node.

Self-loop relation is treated separately. Since each node has exactly one self-edge, we apply a per-node sigmoid, $\alpha^{\text{SELF}}_{ii} = \sigma(e^{\text{SELF}}_{ii})$, producing a node and document-specific gate that determines how much of a node's current representation is retained.

\subsection{Cross-encoder attention}
The embedding of the subject and the embedding of the document are each a single summary vector, so comparing them only tells you about overall topical similarity, not whether the subject matches a specific phrase in the document. To add that finer signal, each node also carries a cross-encoder representation $\mathbf{r}_v$: a vector produced by a fine-tuned cross encoder model that reads the document and the descriptor together and judges how well they match. (These are computed once, ahead of time, and simply looked up during training and inference; if one is missing, it is treated as all-zeros.)

This extra signal is folded into the same attention score, from both the receiving node's and the neighbor's side:
\begin{equation}
e^{r}_{ij}\mathrel{+}=\gamma_r\big(W^{\text{ce}}_t\,\mathbf{r}_i + W^{\text{ce}}_s\,\mathbf{r}_j\big)
\end{equation}
The neighbor's part, $W^{\text{ce}}_s\mathbf{r}_j$ helps the node decide which neighbors to listen to based on how well those neighbors match the document. The strength of this addition is controlled by a single learned number $\gamma_r$ per relation type and is initialized at zero. At the start of training, the layer is therefore identical to the version without the cross-encoder, and the pairwise signal contributes only as $\gamma_r$ grows during training. 

In each layer of message passing on the subgraph, the node's message is now transformed based on the incoming neighbors' message by a learned vector $W^V_r$. The neighbor message is weighed by both the channels, the relation-type edge $\beta$ and the per-neighbor attention
\begin{equation}
\alpha:\mathbf{m}_i = \sum_{r} \beta_{r}\sum_{j\in\mathcal N_r(i)} \alpha^{r}_{ij}\,W^V_r\mathbf{h}_j
\end{equation}
Finally, the node combines this incoming message with its old vector
\begin{equation}
\mathbf{h}'_i = \mathrm{LN}\!\big(\mathbf{h}_i + W_o\,\mathbf{m}_i\big)
\end{equation}
\subsection{Decomposed scoring head}
In addition to the seven interpretable features of Stage-1, the following features are added to each candidate $i$ with refined embedding $\mathbf{h}^1_i$ and initial embedding $\mathbf{h}^0_i$:
\textbf{Semantic match:} $s^{\text{sem}}i = \cos(\mathbf{h}^1_i,\mathbf{d}{\text{doc}})$ which denotes the semantic match of node $i$ (represented by its transformed embedding) with document d. 
\textbf{NT-match (reusing attention)}: For a candidate subject, NT-match identifies how well the narrower terms of the candidate match the document and how well it attends to those that do. 
    \begin{equation}
s^{\text{NT}}_i = \sum_{j\in\mathcal{N}_{\text{NT}}(i)} \alpha^{\text{NT}}_{ij}\,\cos(\mathbf{h}^0_j,\mathbf{d}_{\text{doc}})
\end{equation}
The two parts of the above equation define this match,
\begin{itemize}
    \item \textbf{$\cos(\mathbf{h}^0_j, \mathbf{d}_{\text{doc}})$}: Defines how well the label representation of child $j$ matches the document d. 
    \item  \textbf{$\alpha^{\text{NT}}_{ij}$}: Denotes the attention weights of term $j$ for node $i$ computed in the message passing hidden layer.
\end{itemize} 
\textbf{RT-match}:  $s^{\text{RT}}_i$ is defined in the same way as NT neighbors \textbf{$\text{has\_NT}_i$}, \textbf{$\text{has\_RT}_i \in \{0,1\}$}: Indicates whether a subject candidate contains any narrow or related terms as defined in the thesaurus. This is a binary flag that denotes the missingness of a relation. It is important to distinguish between an edge type that is weak and an edge type that is completely absent in the graph structure.

Although cosine features are easy to understand, they do not provide a sufficient discriminatory signal to separate similar subjects. 
Therefore, in addition to the interpretable feature, we also provide representation directly through an interaction pathway: the vector $[\mathbf{h}^1_i,\; \mathbf{h}^1_i\odot\mathbf{d}_{\text{doc}},\; \mathbf{h}^1_i-\mathbf{d}_{\text{doc}}]$
 is projected to a small hidden vector and concatenated with the rest of the feature. 

\subsection{Prediction}
TAGHN assigns a candidate that exceeds the threshold $\theta$. It is a scalar parameter that is learned along with model training. Since the rules of indexing a document dictate that a more fine-grained subject should be preferred over a general term whenever possible, we apply the BT-suppression filter to any broader-narrower pair in which
both terms have passed $\theta$. By default, it keeps the more specific term: the parent is removed unless it outscores its child by at least
the margin $\delta$, in which case the child is removed instead. A parent
whose child was not predicted is left untouched

\subsection{Training objective}
TAGHN is trained from end-to-end, along with the learned threshold $\theta$ by minimizing a sum of three objectives:
\begin{equation}
\mathcal{L} = \mathcal{L}_{\text{BCE}} + \lambda_1\,\mathcal{L}_{\text{spec}} + \lambda_2\,\mathcal{L}_{\text{NT}}, \qquad \lambda_1 = 0.3,\ \lambda_2 = 0.1
\end{equation}

The first term drives the main relevance decision, whereas the other two encode structural properties specific to thesaurus. All three are evaluated only on assignable candidates, so the context nodes influence neither the gradients nor the learned threshold.

\subsubsection{Relevance term} 
The core objective is a binary cross-entropy over each candidate's threshold-shifted score $\sigma(s_i - \theta)$, where $y_i \in \{0,1\}$ marks whether candidate $i$ is a gold descriptor:
\begin{equation}
\mathcal{L}_{\text{BCE}} = -\frac{1}{|C_d|}\sum_{i}\Big(\gamma_{+}\,y_i\log\sigma(s_i - \theta) + \gamma_{-}\,(1 - y_i)\log\big(1 - \sigma(s_i - \theta)\big)\Big)
\end{equation}

We weight the two error types unequally, $\gamma_{+} = 6$ against $\gamma_{-} = 1.5$. The correct descriptors are heavily outnumbered by the candidates in $C_d$, and under a balanced loss, the model would minimize error simply by predicting almost nothing. Up-weighting the positive term counteracts this imbalance and reflects the cost structure of the task: a missed descriptor is more damaging than a spurious suggestion. Shifting the scores by $\theta$ inside the loss allows the threshold to be learned jointly with the scorer rather than tuned afterward.

Specificity term. Indexers assign the most specific applicable descriptor rather than a broad parent, and the loss enforces this ordering. For each broader–narrower (BT) pair in which the child is a gold label but the parent is not, a hinge penalty pushes the child's probability above the parent's by a margin $m = 0.1$:

\begin{equation}
\mathcal{L}_{\text{spec}} = \operatorname{mean}\;\max\!\big(0,\; m - (\sigma(s_{\text{child}}) - \sigma(s_{\text{parent}}))\big)
\end{equation}

The penalty is zero once the gap exceeds the margin, so the term only acts on pairs that are still ranked incorrectly. This mirrors the specificity preference applied at prediction time, aligning what the model learns with how its output is later filtered.

NT-calibration term. The NT-match feature $s^{\text{NT}}_i$ (§4.4) is read straight from the graph's attention weights, and without supervision, it could drift away from meaning ``the narrower concepts genuinely match.`` A mean-squared-error term ties it back to the gold label:
\begin{equation}
\mathcal{L}_{\text{NT}} = \operatorname{MSE}\big(s^{\text{NT}}_i,\; y_i\big)
\end{equation}
Because $s^{\text{NT}}_i$ is a function of NT attention $\alpha^{\text{NT}}$, this term directly supervises that attention: children of a truly relevant descriptor are pushed to receive weight, and the feature remains a reliable predictor of correctness.

\begin{table*}[t]
\centering
\caption{Stage-1 features cached per candidate}
\label{tab:stage1-features}
\renewcommand{\arraystretch}{1.3}
\begin{tabular}{@{}llp{5.2cm}lc@{}}
\toprule
Field & Model & Computation & Key parameters & Range \\
\midrule
$d_{\text{doc}}$ & BGE-M3 (dense) & Dense encoding of title\,+\,abstract & max\_len 512, dim 1024 & vector \\
\texttt{s\_bgem3} & BGE-M3 (all 3) & $0.4\,\text{dense}+0.2\,\text{sparse}+0.4\,\text{colbert}$ & cand 200, sparse-$k$ 100, top-$k$ 50 & $0$--$0.70$ \\
\texttt{s\_ner} & Llama-3.3-70B + BGE-M3 & Max cosine of NER terms $\rightarrow$ descriptor & ner\_top\_$k$ 3, dim 1024 & $0$--$1.0$ \\
\texttt{in\_sub-thesaurus} & BERT-base & $\sigma(\text{logit})$ of thesaurus node, inherited by its descriptors & top-$k$-thsys 5 & $0$--$0.67$ \\
\texttt{exact\_match} & regex & Whole-word label/altLabel match in text & $\geq 3$ chars & $\{0,1\}$ \\
\texttt{maxsim} & BGE-M3 (ColBERT) & Mean of per-query-token max cosine & label max\_len 64, compact text & $0.22$--$0.76$ \\
\texttt{sparse} & BGE-M3 (sparse) & Dot product over shared tokens & --- & $0$--$0.20$ \\
\bottomrule
\end{tabular}
\end{table*}

\begin{table}[t]
\centering
\caption{Stage-1 candidate generation and recall by source}
\label{tab:stage1-recall}
\renewcommand{\arraystretch}{1.3}
\begin{tabular}{@{}lrrr@{}}
\toprule
Source & Cand./doc & Gold/doc & Recall \\
\midrule
BGE-M3 multivector (\texttt{s\_bgem3})   & $134.07$ & $3.10$ & $0.534$ \\
NER map (\texttt{s\_ner})                & $41.91$  & $1.03$ & $0.178$ \\
Category-derived (\texttt{in\_subthes.}) & $128.38$ & $0.29$ & $0.050$ \\
Direct label cosine                      & $3.68$   & $0.13$ & $0.022$ \\
\midrule
\textbf{Combined pool}                   & $\mathbf{308.03}$ & $\mathbf{4.55}$ & $\mathbf{0.783}$ \\
\bottomrule
\end{tabular}
\end{table}

\section{Experimental setup}

\subsection{Dataset}
We evaluate on a set of intellectually annotated library records, restricted to records that carry both a title and an English-language abstract. The two fields are concatenated to form the document text. The resulting collection contains 13,319 training, 1,664 validation and 1,660 test documents, with 5.80 gold descriptors per document on average and 2,011 distinct subjects appearing in the test gold.

The setting is low-resource relative to the label space. Against 6,198 active subjects, 13319 training documents leave most labels with only a handful of examples: Some of the subjects occur fewer than five times in training, and 70 of the test-gold subjects do not occur at all. Table \ref{tab:metadata-stats} summarizes the metadata and relation coverage available across the vocabulary.

\subsection{Stage-1}
Stage-1 generates possible candidates for a document from four different sources. A BGE-M3 multi-vector embedding model performs semantic matching between dense, sparse and ColBERT representations of the document with dense label embedding. The labels are scored against $0.4,\text{dense} + 0.2,\text{sparse} + 0.4,\text{colbert}$ to generate top 200 candidates.
An NER layer generates domain-specific terms using a Llama-3.3-70B\cite{huang2024llama} model. The faceted nature of indexing (subjects from not just one subject category but spread across different subject category) is incorporated into the prompt, to improve label coverage for each document. The NER terms are then mapped to label embeddings by cosine similarity, keeping only the top three per match. 

A coarse-grained matching utilizes the subject categories mapping of the descriptor/subject. A BERT-base\cite{devlin-etal-2019-bert} model is trained on document text and subject categories mapped to each gold descriptor. A single descriptor can be mapped to one or more subject categories in our thesaurus. The intuition is that this identifies the general topic of the document text, while a downstream reranker can identify the correct descriptor belonging to the category. In addition to semantic matching, we also generate candidates through lexical matching of preferred and alternative label name with the document text. Table \ref{tab:stage1-recall} decomposes the pool by source. The union averages 308 candidates per document and recovers 4.55 of the 5.80 gold descriptors, a recall of 0.783. This is the ceiling for the reranker. Gold descriptors missed here cannot be recovered downstream. The multi-vector retriever contributes most of the recall (0.534), and the remaining sources are individually weak but complementary, which is why we take the union rather than tune a single retriever harder. Raising any source's cutoff flattens quickly while inflating the pool, the last 50 BGE-M3 candidates per document add only 0.033 recall, and the fourth and fifth subject categories add 46 descriptors for 0.012. Fewer than 1 in 150 of these added candidates is a gold descriptor, versus 1 in 68 across the pool, so the cutoffs sit where further candidates enlarge the
pool faster than they recover gold.

\subsection{TAGHN configuration and training}
TAGHN operates at $d = 1024$ with two message-passing layers and dropout $0.1$. We include the top $K_{\text{NT}} = 5$ narrower children and $K_{\text{RT}} = 3$ related terms ranked by cosine similarity to $\mathbf{d}$ as context for each candidate. The model consists of four relation edges: BT, NT, RT (from the thesaurus structure), and SELF carry edges. The scoring head takes 12 features and a 128-dimensional projection of the interaction vector and passes the concatenation through an MLP.

We train for 30 epochs with batch size 32, learning rate $10^{-4}$ and weight decay $3\times10^{-5}$, in mixed precision on Quadro RTX 8000 GPUs. Validation runs every epoch and we early-stop on descriptor F1 with patience 5 and a minimum improvement of 0.001; the reported model stopped at epoch 13. The loss weights are $\lambda_1 = 0.3$, $\lambda_2 = 0.1$

The positive and negative weights in the relevance term $\gamma_+/\gamma_- = 6.0/1.5$ were selected based on the validation set. The ratio controls where the model sits on the precision–recall trade-off.

During inference, the learned threshold is used directly ($\theta = -0.053$), followed by BT-suppression with margin $\delta = 0.05$. TAGHN predicts 5.64 descriptors per document on average versus 5.80 gold.

TAGHN operates on document-specific subgraphs whose size is bounded by the candidate budget (typically 230–420 nodes per document in STW), its per document cost does not grow with thesaurus size. Earlier graph-based methods propagate over the full label graph at every step and keep all label embeddings in GPU memory, so both compute and memory grow with the vocabulary. In practice, the full model (57M parameters, fp16, precomputed embeddings) trained on 13.3k STW documents for 30 epochs with early stopping on a single Quadro RTX 8000, including validation after every epoch. Scoring all 1,660 test documents took less than 4 minutes end to end, or less than 0.15s per document.

\subsection{Baseline models}
TAGHN is evaluated against two types of model, retriever based models that uses the same retrieve-rerank pipeline and models that are designed for XMLC that assume a fixed vocabulary. 
The retrieve-rerank models share the same candidate pool as TAGHN from Stage-1. However, the fixed vocabulary model does not require a candidate pool and works directly on the entire descriptor list.

The fixed vocabulary baseline models are Omikuji-Bonsai and the XR-Tansformer. Omikuji is an implementation of partitioned label tree methods with Parabel and Bonsai variants, which are integrated as one of the backends in the Annif tool. We train it through annif in the Bonsai configuration using snowball analyzer, bigram features, and model specific parameter such as maximum depth, number of cluster.
The XRtransformer is available in the pecos\cite{yu2022pecos} bundle. Its label space is clustered by spherical hierarchical $k$-means, the matcher is trained for 10 epochs with batch size 32, learning rate $5\times10^{-5}$ and input truncation at 256 tokens, using teacher-forced negatives and a squared-hinge loss, and the linear ranker is fitted over the resulting shortlists. We select the value of Cn and Cp based on the validation set. Both classifiers are trained on the same $13,319$ documents and predict using the full vocabulary.

Among the retriever based methods, we evaluate a training-free zero-shot model that uses BGE-m3 embedding model at its core. We employ \texttt{BAAI/bge-reranker-v2-m3}\cite{chen2024bge} off the shelf with no task adaptation; it establishes how much of the task is solved by a general-purpose reranker and how far the vocabulary is from its pretraining distribution.
The \emph{trained cross-encoder} fine-tunes the same checkpoint on our $13,319$ training documents for 10 epochs with batch size 32, learning rate $2\times10^{-5}$, linear warm-up over the first 10\% of steps, weight decay $0.01$ and maximum sequence length 512, with early stopping.
The descriptor side is the bare preferred label rather than the enriched label text used for retrieval; the enriched form was tried first and performed worse, as the concatenated aliases and scope notes dilute the pair signal the cross-encoder relies on. The decision threshold is selected on the validation split.

\textsc{TAGHN}~(w/o CE) is our own system with the cross-encoder attention channel removed (\S4.3), reported along with the baselines to separate the contribution of the graph from that of the pairwise signal folded into it.

All systems are trained on the same $13,319$ documents, evaluated on the same $1,660$ test documents against the same active vocabulary of 6,198 descriptors, and scored with a single evaluation script.

\subsection{Evaluation metrics}
We report precision, recall and F1-score aggregated by different means to capture the highly skewed distribution of subjects. 
\begin{itemize}
    \item \textbf{Micro:} Pools every document, subject before aggregating. Since every prediction counts, it is dominated by head labels
    \item \textbf{Macro:} Computes precision, recall and F1-scores for each subject and averages over them. It is a tail-sensitive view and provides performance across label frequency.
\end{itemize}

\section{Results and discussion}

The table \ref{tab:main-results} reports the performance on the test set according to different aggregated metrics. Significance is assessed by a paired approximate randomization test with Holm-correction applied to the table.

The TAGHN model outperforms all other systems in three out of six metrics namely macro precision, recall and F1 and each of these are statistically significant with every baseline. On micro F1-score, TAGHN reaches 0.487 which is in close competition with the XRtransformer. Against the strongest reranker baseline, a fine-tuned bge-m3 cross encoder, the TAGHN model (without incorporating the cross-encoder) scores a +4.6 increase in micro F1-score while maintaining the macro F1-scores. Combining the complementary signal from the cross-encoder, we are able to show a significant increase of +4.4 points on micro F1 and +3.3 macro F1-score over TAGHN without the cross-encoder. Similarly, we can observe an improvement of +9 micro F1 points and +3.4 macro F1 points over fine-tuned bge-m3 cross-encoder model.

Compared to XMLC specialised model such as bonsai, TAGHN adds a +9.3 micro F1 and +16 macro F1 points. In terms of macro F1, TAGHN is ahead of the XRTransformer by +10.2 points. However it falls behind in micro F1 by 3.0 points primarily due to micro F1 being dominated by head labels.

% Requires \usepackage{booktabs}
\begin{table}[t]
  \centering
  \caption{Test-set subject indexing performance under micro and macro aggregation}
  \small
  \setlength{\tabcolsep}{4pt}
  \begin{tabular}{@{}l ccc ccc@{}}
    \toprule
    & \multicolumn{3}{c}{Micro} & \multicolumn{3}{c}{Macro (label-avg.)} \\
    \cmidrule(lr){2-4}\cmidrule(lr){5-7}
    Model & P & R & F$_1$ & P & R & F$_1$ \\
    \midrule
    Zero-shot (bge-reranker)   & 0.193 & 0.221 & 0.206 & 0.262 & 0.241 & 0.217 \\
    TAGHN (w/o CE)         & 0.415 & 0.475* & 0.443 & 0.352 & 0.392 & 0.347 \\
    CE-trained      & 0.376 & 0.419 & 0.397 & 0.358 & 0.390 & 0.346 \\
    \textbf{\textsc{TAGHN} (ours)} & 0.494 & 0.481 & 0.487 & \textbf{0.398} & \textbf{0.415} & \textbf{0.380} \\
    \midrule
    Omikuji-Bonsai & 0.303 & \textbf{0.563*} & 0.394 & 0.224 & 0.280 & 0.220 \\
    \textsc{Xrt} (PECOS)       & \textbf{0.584} & 0.463 & \textbf{0.517} & 0.338 & 0.263 & 0.278 \\
    \bottomrule
  \end{tabular}
  \label{tab:main-results}
  \begin{minipage}{\columnwidth}
    \footnotesize
    $^{*}$Difference from \textsc{TAGHN} not significant (paired approximate
    randomization test, Holm-corrected per table, $\alpha = 0.05$).
  \end{minipage}
\end{table}

\subsection{Performance across label frequency}
\begin{table*}[t]
  \centering
  \caption{Macro-averaged F1 by training-frequency bin}
  \small
  \setlength{\tabcolsep}{5pt}
  \begin{tabular}{@{}lrcccccc@{}}
    \toprule
    Train frequency & \#lbl & \textsc{TAGHN (w/o CE)} & Zero-shot (bge-reranker) & CE-trained & \textsc{TAGHN}\ & Omikuji & \textsc{Xrt} \\
    \midrule
    $0$ (unseen)  & 70   & 0.086* & 0.195* & \textbf{0.288} & 0.114 & 0.000 & 0.000 \\
    $1$--$5$      & 432  & 0.257* & 0.191 & \textbf{0.286*} & \textbf{0.278} & 0.040 & 0.095 \\
    $6$--$20$     & 728  & 0.342 & 0.209 & 0.337 & \textbf{0.388} & 0.159 & 0.235 \\
    $21$--$100$   & 663  & 0.410 & 0.233 & 0.379 & \textbf{0.438} & 0.333 & 0.420* \\
    $101$--$500$  & 110  & 0.502 & 0.284 & 0.457 & 0.540 & 0.406 & \textbf{0.571} \\
    $>500$        & 8    & 0.467* & 0.187* & 0.404* & 0.464 & 0.455* & \textbf{0.660}* \\
    \midrule
    All           &  & 0.347 & 0.217 & 0.346 & \textbf{0.380} & 0.220 & 0.278 \\
    \bottomrule
    
  \end{tabular}
  \label{tab:tail-label-F1}
\begin{minipage}{\columnwidth}
    \footnotesize
    $^{*}$Difference from \textsc{TAGHN} not significant (paired approximate
    randomization test, Holm-corrected per table, $\alpha = 0.05$).
  \end{minipage}
\end{table*}

The table \ref{tab:tail-label-F1} shows how the performance of various models is distributed across subject frequency. We report the test-set labels binned by their frequency on training dataset. Consistent performance across label frequency is an essential requirement of the subject indexing, because the task is often characterized by long tails. 

TAGHN leads the mid-frequency bins, reaching 0.388 on 6–20 (728 labels) and 0.438 on 21–100 (663 labels), and it attains the best aggregate macro-F1 of 0.380 over all 2,011 test-gold descriptors ahead of the graph reranker without the cross-encoder (0.347), the trained cross-encoder (0.346), Xrt (0.278), and Omikuji (0.220). The Xrt is significantly stronger on frequent labels, taking 101–500 (0.571 vs. 0.540). It also leads above 500 occurrences (0.660 vs. 0.464, p=0.009), but this bin contains only eight descriptors, which limits how small a p-value the randomization test can produce. It also degrades sharply into the tail and scores 0.000 on unseen labels by construction, having no parameters for descriptors absent from training. TAGHN  concentrates its advantage on the 6–100 range, which covers 1,391 descriptors against only 118 above 100, the region where the bulk of real indexing work happens. 

On the 70 unseen descriptors, trained cross-encoder (0.288) remains the strongest system and is significantly better than TAGHN (0.114). While, zero-shot bge-m3 model (0.195, p=0.124) scores better than TAGHN, it is not statistically significant. On the 1–5 bin, the difference between the trained cross-encoder and TAGHN is not significant (0.286 vs. 0.278, p=0.705), and we regard the two systems as tied. 
The configuration that maximizes aggregate F1 is thus not the configuration that best serves the extreme tail.

\subsection{Head-to-head comparison}
Aggregate metrics can conceal whether a system wins broadly or wins by a big margin on a few documents. Comparing TAGHN and the trained cross-encoder document-by-document on the 1,660 documents both systems index, TAGHN achieves a higher per-document F1 on 1,100 documents (66.3\%), a lower one on 415 (25.0\%), and ties on 145 (8.7\%). Mean per-document F1 is 0.483 versus 0.393. The number of documents for which a system retrieves nothing correct - the failure mode most visible to a human indexer - falls from 79 (4.8\%) to 59 (3.6\%). The advantage is therefore distributional rather than driven by outliers, though the 25\% of documents on which the cross-encoder remains better indicates the two signals are complementary rather than redundant.

\subsection{Qualitative analysis}
\subsubsection{Recovering a coherent concept family}
Looking at a specific document titled ``\textit{Structural separation and access in telecommunications markets}`` annotated with five subjects: \textit{Access regulation, Spin-off company, Telecommunications, Telecommunications policy, and Theory}. The trained cross-encoder identifies none of them. It returns \textit{Bottleneck, Competition, Local area network, Market structure, Regulation, Segregation,} and \textit{Structural break}. All of these subjects taken individually are adequate for a document about regulated network access, but none of them is the one chosen by subject indexers. It proposes a \textit{Regulation} where the gold label is the narrower \textit{Access regulation}, and \textit{Local area network}, where the topic is telecommunications infrastructure generally. TAGHN identifies four of the five (Access regulation, Telecommunications, Telecommunications policy, Theory), missing only the \textit{Spin-off company}. The cross-encoder scores each subject against the document independently and thus cannot identify that Regulation has a narrower term that fits better, whereas message passing over the BT/NT edges lets the parent's score be informed by its children's document match. 
While both the models make errors, the errors are not too far off from the correct labels. For example, TAGHN also yields the \textit{Bottleneck} and \textit{Telecommunications network}. However, these false positives are near-misses within the correct region of the thesaurus. 
\subsubsection{Edge gates at a rare leaf}
Document with title `\textit{A review of the German
mandatory deposit for one-way drinks packaging and drinks packaging taxes in
Europe}' contains the gold descriptor \emph{Drinks packaging}, which is seen
exactly \textbf{once} in the training set and therefore falls in the $1$--$5$
bin of Table~\ref{tab:tail-label-F1}. It was identified successfully by Stage-1 and was part of the result set by the reranker.

In the thesaurus, \emph{Drinks packaging} is a leaf node: it has no narrower terms. The edges in the per-document subgraph, weighted by the effective edge
weight $\beta_r \cdot \alpha^{r}_{ij}$, are: 
(Channel-A gates at layer 1: BT $0.318$, NT $0.473$, RT $0.451$, SELF $0.558$.) Since it is a leaf node, its NT-match feature is undefined. 

\begin{table}[t]
  \centering
  \caption{Effective edge weights for the rare leaf descriptor \emph{Drinks packaging}}
  \small
  \setlength{\tabcolsep}{5pt}
  \begin{tabular}{@{}llcc@{}}
    \toprule
    Relation & Neighbor (train freq.) & $\alpha$ & $\beta_r\!\cdot\!\alpha$ \\
    \midrule
    BT   & Food packaging (0)       & 1.000 & 0.318 \\
    RT   & Reusable packaging (0)   & 0.616 & 0.278 \\
    RT   & Disposable packaging (0) & 0.384 & 0.173 \\
    RT   & Metal packaging (0)      & 0.000 & 0.000 \\
    SELF & (itself)                 & $\approx$0.000 & $\approx$0.000 \\
    NT   & --- none ---             & ---   & ---   \\
    \bottomrule
  \end{tabular}
  \label{tab:edge-weight}
\end{table}

The has\_NT missingness indicator was designed to help the scoring head  distinguish between ``no children`` and ``children mismatch``. Thus, exposing the fact that the subject is a leaf node to the score head. Since NT information is unavailable, the model falls back to RT context, concentrating on \textit{Reusable packaging} and \textit{Disposable packaging} while driving \textit{Metal packaging} to zero. It should be noted that all three RT neighbors and the BT parent are not seen in training themselves. The graph supplies tail-to-tail structural context, relating an already observed subject to its neighbors, which are previously unseen. A case that would be challenging for a frequency-dependent classifier is handled well with the help of the graph structure. In table \ref{tab:edge-weight} The BT edge carries the largest weight (0.318) but is capped by a low $\beta_BT$; consistent with the specificity objective, the broad parent does not dominate the signal. We can observe that the SELF gate collapses to $\approx!0$ ($0.0003$, the lowest of any node in this document, against a corpus median of $0.70$): the refined representation is built almost entirely from neighbors rather than from the subject itself, and correspondingly its graph-derived semantic score is weak ($s_{\text{sem}} = 0.158$). What carries the label instead are the signals from Stage-1 that bypass message - passing a verbatim surface match ($\text{exact} = 1.0$), late interaction ($s_{\text{maxsim}} = 0.694$), and Stage-1 dense similarity ($s_{\text{bgem3}} = 0.592$) - together with RT context ($s_{\text{RT}} = 0.535$). The graph situates the descriptor; the lexical and dense signals assign it.

When comparing the trained cross-encoder, it also recovers
\emph{Drinks packaging} on this document, and does so from five predictions to
TAGHN's seven. In this document, TAGHN attains a per-document $F_1$ of
$0.286$ against the cross-encoder's $0.167$, but both models index it poorly.

\subsection{Comparison with fixed-vocabulary classification}
The Omikuji-Bonsai attains the highest recall in Table \ref{tab:main-results} (0.563 micro). However, its increase in recall is mainly due to predicting substantially more descriptors per document, and its precision (0.303) and macro F1 (0.220) are correspondingly the weakest among all the systems. 

The XRtransformer model outperforms all other models in two specific bins, namely, 101-500, >500, which only includes 5.9\% of the test label set. Its performance drops sharply on descriptors seen fewer than 20 times. This is a strong indicator that the model caters strongly to the head label.

In Table \ref{tab:tail-label-F1}, all retrieval-based systems, including the training-free cross-encoder, recover unseen labels to some degree because the label space is open by construction. But Omikuji and the XRtransformer score 0.000 on the unseen descriptor because a fixed-vocabulary classifier has no training signal for a label absent from its training set. It also collapses in the rare bins (1–5,6–20) and only becomes competitive above 500 and 100 frequency bins, respectively. However, this subset contains only a small portion of 2,011 labels. 

\section{Conclusion and future works}

We present TAGHN, a heterogeneous graph reranker for automated subject indexing that treats the controlled vocabulary as a structured graph rather than a flat label list. TAGHN scores each candidate over a per-document subgraph of Stage-1 candidates augmented with non-assignable BT/NT/RT context nodes, conditioning every part of message passing on the document through a per-relation edge-type gate and a three-way attention over target, neighbor, and document. Cross-encoder representations enter the same attention behind a gate initialized at zero. A decomposed scoring head combines graph-derived features with Stage-1 retrieval signals.

% On STW, an economic thesaurus of 6,198 descriptors under a low-resource regime, TAGHN reaches 0.380 macro F1, ahead of a fine-tuned cross-encoder (0.346), XR-Transformer (0.278), and Omikuji-Bonsai (0.220), while remaining competitive on micro F1 (0.487 vs. 0.517). The improvement is precision at constant recall: TAGHN emits 5.65 descriptors per document to the cross-encoder's 6.46, with true and false positives at a near 1:1 ratio (4,645 to 4,759) against the cross-encoder's 4,039 to 6,692. Holding recall while predicting fewer labels means gold descriptors genuinely moved up the ranking. TAGHN wins per-document F1 on 66.3\% of documents and cuts the indexer-visible failure of returning nothing correct from 4.8\% to 3.6\% compared to the cross-encoder. The per-frequency view shows that TAGHN and XR-Transformer are mirror images, the latter strongest above 500 occurrences (0.660 vs. 0.464) and zero on unseen labels by construction, while TAGHN concentrates its advantage on the 6–100 range containing 1,391 of 2,011 test-gold descriptors.

On STW, an economics thesaurus of 6,198 descriptors under a low-resource regime, TAGHN reaches 0.380 macro F1, ahead of a fine-tuned cross-encoder (0.346), the XRTransformer (0.278), and Omikuji-Bonsai (0.220), while remaining competitive on micro F1 (0.487 vs. 0.517). The gain over the cross-encoder comes mainly from precision. TAGHN emits 5.65 descriptors per document against the cross-encoder's 6.46, yet recovers more gold descriptors (4,645 vs. 4,039 true positives), with true and false positives at a near 1:1 ratio (4,645 to 4,759) against the cross-encoder's 4,039 to 6,692. Predicting fewer labels while recovering more of them indicates that gold descriptors moved up the ranking. TAGHN wins per-document F1 in 66.3\% of documents and reduces the indexer-visible failure to return nothing correct from 4.8\% to 3.6\%. The per-frequency view shows that TAGHN and the XRTransformer are near mirror images. The XRTransformer is strongest above 100 occurrences and scores zero on unseen labels by construction, while TAGHN concentrates its advantage in the 6 to 100 range, which contains $1,391$ of the $2,011$ test-gold descriptors.

For future work, we would like to focus our efforts on raising the Stage-1 ceiling. It gives us the highest leverage, since 0.783 caps every downstream gain. We also plan to validate TAGHN on other vocabularies, at label scales of 20,000+ descriptors, and in the cross-lingual setting that our bilingual thesaurus supports but our English-only evaluation does not exercise. In addition, a more detailed ablation study, isolating the edge-type gate, neighbor attention, context nodes, and structural loss terms, remains to be carried out; the present evaluation removes only the cross-encoder channel. Finally, we will explore enriching the neighborhood through cross-vocabulary mappings. The relation edges among the thesaurus structure are unevenly distributed. Of the 6,198 active descriptors, from the table \ref{tab:metadata-stats}, only 28.3\% have a narrower term. Tail descriptors are specific by nature and therefore sit at the leaves of the hierarchy with no children at all. This suggests that importing neighborhood structure could be useful for tail labels. Because the thesaurus is published as linked open data in SKOS, its descriptors can be aligned to concepts in other vocabularies through skos:exactMatch and skos:closeMatch links, and the neighborhoods those vocabularies provide can be transplanted onto the local graph.
%%
% The acknowledgments section is defined using the "acks" environment
% (and NOT an unnumbered section). This ensures the proper
% identification of the section in the article metadata, and the
% consistent spelling of the heading.
% \begin{acks}
% I am grateful to Prof.~Dr.~Ralf Krestel (ZBW -- Leibniz Information Centre for
% Economics and Kiel University) for supervising this work and for his
% substantial contributions to it. His guidance
% shaped the direction of the paper, and his feedback on the design of the
% reranker, the evaluation, and earlier drafts substantially improved it. I also thank ~Dr. Argie Kasprzik for supporting this research at ZBW.
% \end{acks}

\section*{Ethics and privacy statement}
This work uses bibliographic metadata describing published works, not people, and
involves no human subjects. The system is intended to propose descriptors for a
professional indexer to accept or reject, not to index unattended: its accuracy is uneven across the vocabulary, weakest on the unseen descriptors, so unsupervised use would systematically under-apply the specific terms that make a collection navigable.

% \printbibliography
\bibliographystyle{ACM-Reference-Format}
\bibliography{references}

\

\end{document}